\documentclass[runningheads]{llncs}
\usepackage{graphicx}
\usepackage[utf8]{inputenc} % allow utf-8 input
\usepackage[T1]{fontenc}    % use 8-bit T1 fonts
\usepackage{hyperref}       % hyperlinks
\usepackage{url}            % simple URL typesetting
\usepackage{booktabs}       % professional-quality tables
\usepackage{amsfonts}       % blackboard math symbols
\usepackage{nicefrac}       % compact symbols for 1/2, etc.
\usepackage{microtype}      % microtypography

\usepackage{hyperref}

\usepackage{url}
\usepackage{algorithm}
\usepackage{algorithmic}
\usepackage{graphicx}
\usepackage{amssymb}
\usepackage{multicol}
\usepackage{amsmath}
\usepackage{array}

\newcolumntype{x}[1]{%
>{\centering\hspace{0pt}}p{#1}}%

\usepackage{float}
\usepackage{bbm,ifthen,twoopt,color}

\usepackage[utf8]{inputenc} % allow utf-8 input
\usepackage[T1]{fontenc}    % use 8-bit T1 fonts
\usepackage{hyperref}       % hyperlinks
\usepackage{url}            % simple URL typesetting
\usepackage{booktabs}       % professional-quality tables
\usepackage{amsfonts}       % blackboard math symbols
\usepackage{nicefrac}       % compact symbols for 1/2, etc.
\usepackage{microtype}      % microtypography
\usepackage{dsfont}
\usepackage{graphicx}
\usepackage{subcaption}
\usepackage{adjustbox}

\usepackage{soul}
\usepackage{footmisc}
\usepackage{url}
\usepackage{wrapfig}
\usepackage{subcaption}
\usepackage{multirow}
\usepackage{multicol}
\usepackage{tabularx}
\usepackage{adjustbox}
\begin{document}
\title{Online Learning in Iterated Prisoner's Dilemma to Mimic Human Behavior
}
\titlerunning{Online Learning in Iterated Prisoner's Dilemma to Mimic Human Behavior
}

\author{Baihan Lin\inst{1,*}, Djallel Bouneffouf\inst{2} \and  Guillermo Cecchi\inst{2}}
\authorrunning{B. Lin, D. Bouneffouf and G. Cecchi}
\institute{Columbia University, New York, USA \and
IBM Research, Yorktown Heights, NY, USA \\
\email{baihan.lin@columbia.edu,  djallel.bouneffouf@ibm.com, gcecchi@us.ibm.com}\\}

\maketitle

\begin{abstract}

As an important psychological and social experiment, the Iterated Prisoner’s Dilemma (IPD) treats the choice to cooperate or defect as an atomic action. We propose to study the behaviors of online learning algorithms in the Iterated Prisoner’s Dilemma (IPD) game, where we investigate the full spectrum of reinforcement learning agents: multi-armed bandits, contextual bandits and reinforcement learning. We evaluate them based on a tournament of iterated prisoner's dilemma where multiple agents can compete in a sequential fashion. This allows us to analyze the dynamics of policies learned by multiple self-interested independent reward-driven agents, and also allows us study the capacity of these algorithms to fit the human behaviors. Results suggest that considering the current situation to make decision is the worst in this kind of social dilemma game. Multiples discoveries on online learning behaviors and clinical validations are stated, as an effort to connect artificial intelligence algorithms with human behaviors and their abnormal states in neuropsychiatric conditions. 
% \footnote{The data and codes to reproduce all the empirical results can be accessed at \href{https://app.box.com/s/ytcz067bjv1dhy82d7ho7afu7v3pprs5}{\underline{https://app.box.com/s/ytcz067bjv1dhy82d7ho7afu7v3pprs5}}.}
\footnote{The data and codes to reproduce all the empirical results can be accessed at \href{https://github.com/doerlbh/dilemmaRL}{\underline{https://github.com/doerlbh/dilemmaRL}}.}
\end{abstract}


\section{Introduction}

Social dilemmas expose tensions between cooperation and defection. 
Understanding the best way of playing the iterated prisoner's dilemma (IPD) has been of interest to the scientific community since the formulation of the game seventy years ago \cite{axelrod1980effective}. To evaluate the algorithm a round robin computer tournament was proposed, where algorithms competed against each others \cite{andreoni1993rational}. The winner was decided on the average score a strategy achieved. Using this framework, we propose here to focus on studying reward driven online learning algorithm with different type of attentions mechanism,
where we define attention "as the behavioral and cognitive process of selectively concentrating on a discrete stimulus while ignoring other perceivable stimuli" \cite{johnson2004attention}. 
 Following this definition, we analyze three algorithms classes: the no-attention-to-the-context online learning agent (the multi armed bandit algorithms) outputs an action but does not use any information about the state of the environment (context); the contextual bandit algorithm extends the model by making the decision conditional on the current state of the environment, and finally reinforcement learning as an extension of contextual bandits which makes decision conditional on the current state of the environment and the next state of the unknown environments. 
 This paper mainly focuses on an answer to two questions:
\begin{itemize} 

    \item Does attending to the context for an online learning algorithm helps on the task of maximizing the rewards in an IPD tournament, and how do different attention biases shape behavior?

    \item Does attending to the context for an online learning algorithm helps to mimic human behavior?

\end{itemize}
 To answer these questions, we have performed two experimenters:
(1) The first one where we have run a tournament of the iterated prisoner's dilemma: Since the seminal tournament in 1980 \cite{axelrod1980effective}, a number of IPD tournaments have been undertaken \cite{andreoni1993rational,bo2005cooperation,bereby2006speed}.
% ,duffy2009cooperative,kunreuther2009bayesian,dal2011evolution,friedman2012continuous,fudenberg2012slow,harper2017reinforcement}. 
In this work, we adopt a similar tournament setting, but also extended it to cases with more than two players. Empirically, we evaluated the algorithms in four settings of the Iterated Prisoner's Dilemma: pairwise-agent tournament, three-agent tournament, ``mental''-agent tournament.
(2) Behavioral cloning prediction task: where we train the the three types of algorithm to mimic the human behavior on some training set and then test them in a test set. 
Our main results are the following:

\begin{itemize} 

    \item We observe that contextual bandits are not performing well in the tournament, which means that considering the current situation to make decision is the worst in this kind of social dilemma game. Basically we should either do not care about the current situation or caring about more situations, but not just the current one.
    
    \item We observe that bandit algorithms (without context) is the best in term of fitting the human data, which implies that humans may not consider the context when they play the iterated prisoner's dilemma.

    \end{itemize}

This paper is organized as follows. We first review related works and introduces some background concepts. Then we explain the two experiments we have performed. Experimental evaluation highlights the empirical results we have got. Finally, the last section concludes the paper and points out possible directions for future works. 

As far as we are aware, this is the first work that evaluated the online learning algorithms in social gaming settings. Although the agents that we evaluated here are not newly proposed by us, we believe that given this understudied information asymmetry problem setting, our work helps the community understand how the inductive bias of different methods yield different behaviors in social agent settings (e.g. iterated prisoners' dilemma), and thus provides a nontrivial contribution to the fields, both in understanding machine learning algorithms, and in studying mechanistic models of human behaviors in social settings.

\section{Related Work}

There is much computational work focused on non understanding the strategy space and finding winning strategies in the iterated prisoner's dilemma; Authors in \cite{kies2020finding} present and discuss several improvements to the Q-Learning algorithm, allowing for an easy numerical measure of the exploitability of a given strategy. \cite{Gaurav2020} propose a mechanism for achieving cooperation and communication in Multi-Agent Reinforcement Learning settings by intrinsically rewarding agents for obeying the commands of other agents. We are interested in investigating how algorithms are behaving and also how they are modeling the human decisions in the IPD, with the larger goal of understanding human decision-making. For-instance, In \cite{park2016active} authors have proposed an active modeling technique to predict the behavior of IPD players. The proposed method can model the opponent player’s behavior while taking advantage of interactive game environments. The data showed that the observer was able to build, through direct actions, a more accurate model of an opponent’s behavior than when the data were collected through random actions. \cite{capraro2013model} they propose the first predictive model of human cooperation able to organize a number of different experimental findings that are not explained by the standard model and they show also that the model makes satisfactorily accurate quantitative predictions of population average behavior in one-shot social dilemmas. To the best of our knowledge no study has been exploring the full spectrum of reinforcement learning agents: multi-armed bandits, contextual bandits and reinforcement learning in social dilemma.
 
% \newpage

\section{Background}
\label{sec:problem}

% Here we briefly outline the three online learning algorithms: 

\textbf{Multi-Armed Bandit (MAB):} The multi-armed bandit (MAB) algorithm models a sequential decision-making process, where at each time point a the algorithm selects an action from a given finite set of possible actions, attempting to maximize the cumulative reward over time \cite{LR85,UCB}. 
 
%Optimal solutions have been provided using a stochastic formulation , or using an adversarial formulation \cite{AuerC98,AuerCFS02,BouneffoufF16}.
%Recently, there has been a surge of interest in a Bayesian formulation \cite {chapelle2011empirical}, involving the algorithm known as Thompson sampling \cite {T33}. Theoretical analysis in \cite{AgrawalG12} shows that Thompson sampling for Bernoulli bandits is asymptotically optimal. \\
% asymptotically achieves the optimal performance limit. \\
% Lai and Robbins (1985)
% Empirical analysis of Thompson sampling, including problems more complex than the Bernoulli bandit, demonstrates that its performance is highly competitive with other approaches \cite{chapelle2011empirical,DBouneffouf14}.

% \subsection{Contextual Bandit (CB)}

\textbf{Contextual Bandit  Algorithm (CB).}
Following \cite{langford2008epoch}, this problem is defined as follows. At each time point (iteration) $t \in \{1,...,T\}$, an agent is presented with a {\em context} ({\em feature vector}) $\textbf{x}_t \in \mathbf{R}^N$
  before choosing an arm $k  \in A = \{ 1,...,K\} $.
We will denote by
  $X=\{X_1,...,X_N\}$  the set of features (variables) defining the context.
Let ${\textbf r_t} = (r^{1}_t,...,$ $r^{K}_t)$ denote  a reward vector, where $r^k_t \in [0,1]$ is a reward at time $t$  associated with the arm $k\in A$.
Herein, we will primarily focus on the Bernoulli bandit with binary reward, i.e. $r^k_t \in \{0,1\}$.
Let $\pi: X \rightarrow A$ denote a policy.  Also, $D_{c,r}$ denotes a joint distribution over  $({\bf x},{\bf r})$.
%Following \cite{ChuLRS11},
We will assume that the expected reward is a linear function of the context, i.e.
%\begin{eqnarray*}
$E[r^k_t|\textbf{x}_t] $ $= \mu_k^T \textbf{x}_t$,
%\end{eqnarray*}
where $\mu_k$ is an unknown weight vector associated with arm $k$. 
%\begin{definition}

% \subsection{Reinforcement Learning}	\label{subsec:rl}
\textbf{Reinforcement Learning (RL).}
Reinforcement learning defines a class of algorithms for solving problems modeled as Markov decision processes (MDP) \cite{Sutton1998}. An MDP is defined by the tuple $(\mathcal{S}, \mathcal{A}, \mathcal{T}, \mathcal{R}, \gamma)$, where  $\mathcal{S}$ is a set of possible states, $\mathcal{A}$ is a set of actions, $\mathcal{T}$ is a transition function defined as $\mathcal{T}(s, a, s')=\Pr(s'\vert s,a)$, where $s, s'\in \mathcal{S}$ and $a\in \mathcal{A}$, and $\mathcal{R}: \mathcal{S}\times \mathcal{A} \times \mathcal{S}\mapsto \mathbb{R}$ is a reward function, $\gamma$ is a discount factor that decreases the impact of the past reward on current action choice. Typically,  the objective is to maximize the discounted long-term reward, assuming  an infinite-horizon decision process, i.e. to find a policy function $\pi: \mathcal{S} \mapsto \mathcal{A}$ which specifies the action to take in a  given state, so that the cumulative reward is maximized: $\max_{\pi} \sum_{t=0}^{\infty}\gamma^t \mathcal{R}(s_t,a_t, s_{t+1}).
$

\section{Experimental Setup}

\begin{figure*}[tb]

\centering
\includegraphics[width=0.24\linewidth]{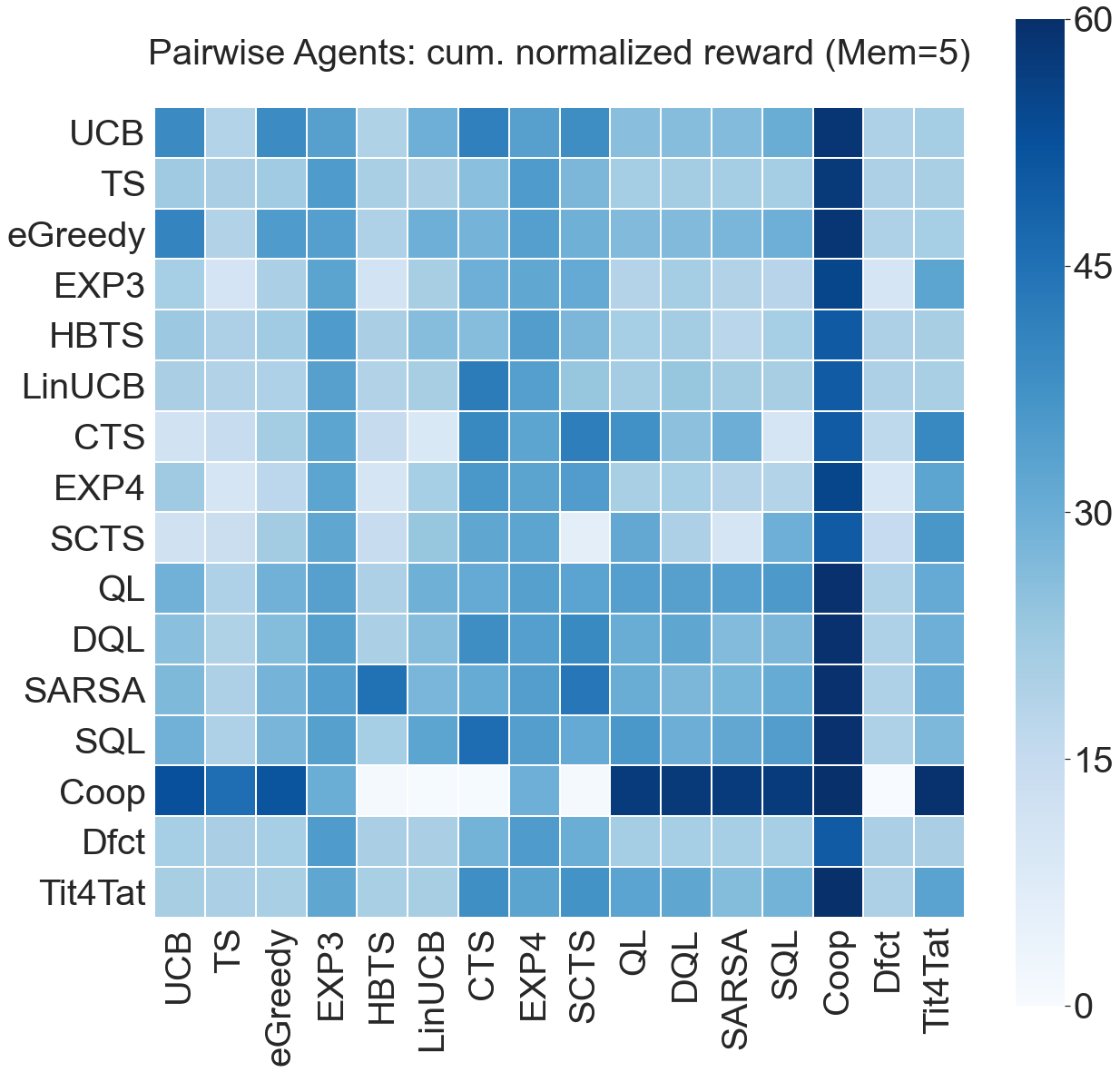}\hfill
\includegraphics[width=0.24\linewidth]{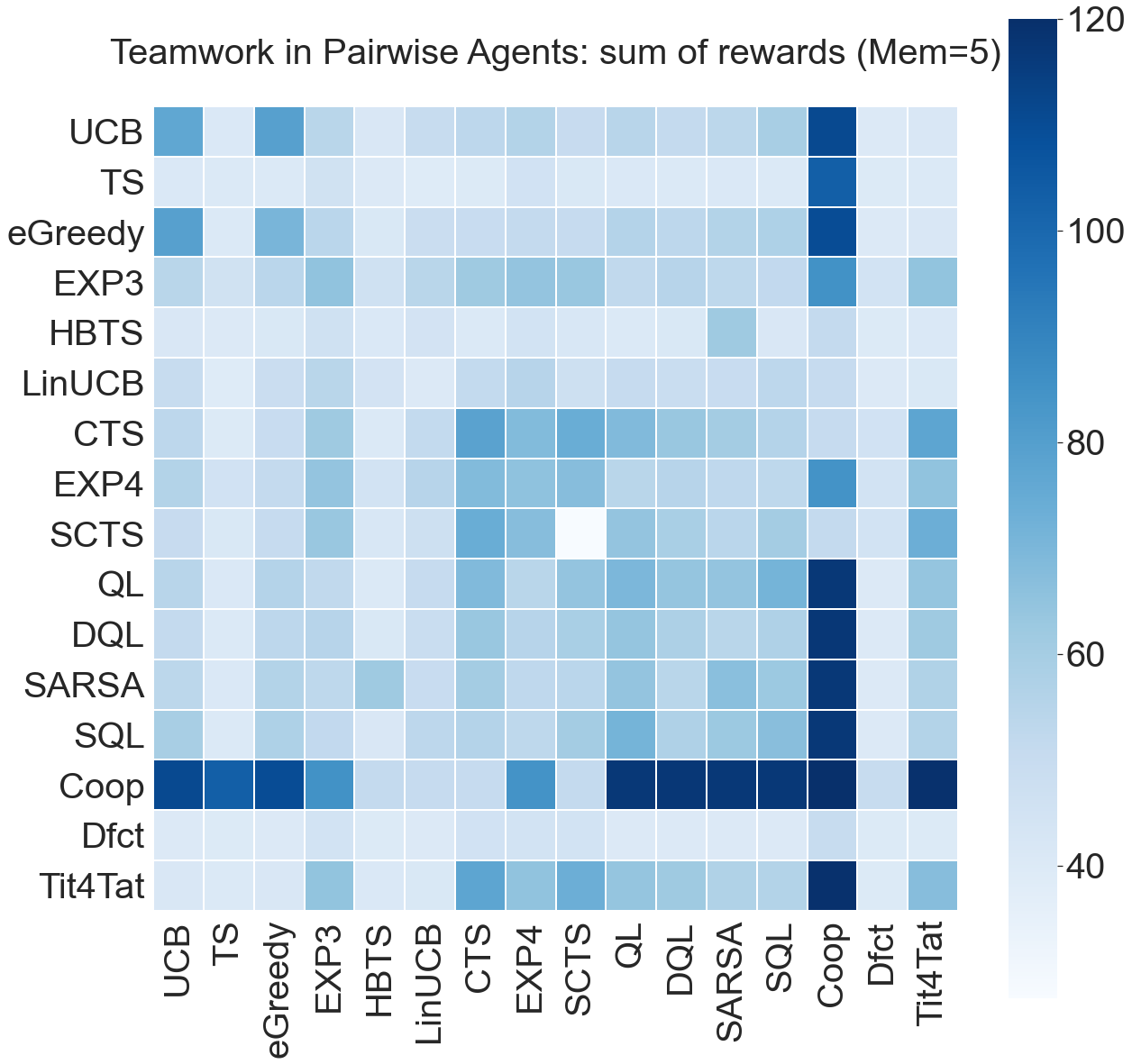}\hfill
% \par\caption{Success and teamwork in two-agent tournament: individual rewards and collective rewards.}\label{fig:rp}
% \end{figure}
% \begin{figure}[h!]
% % 
% \centering
\includegraphics[width=0.24\linewidth]{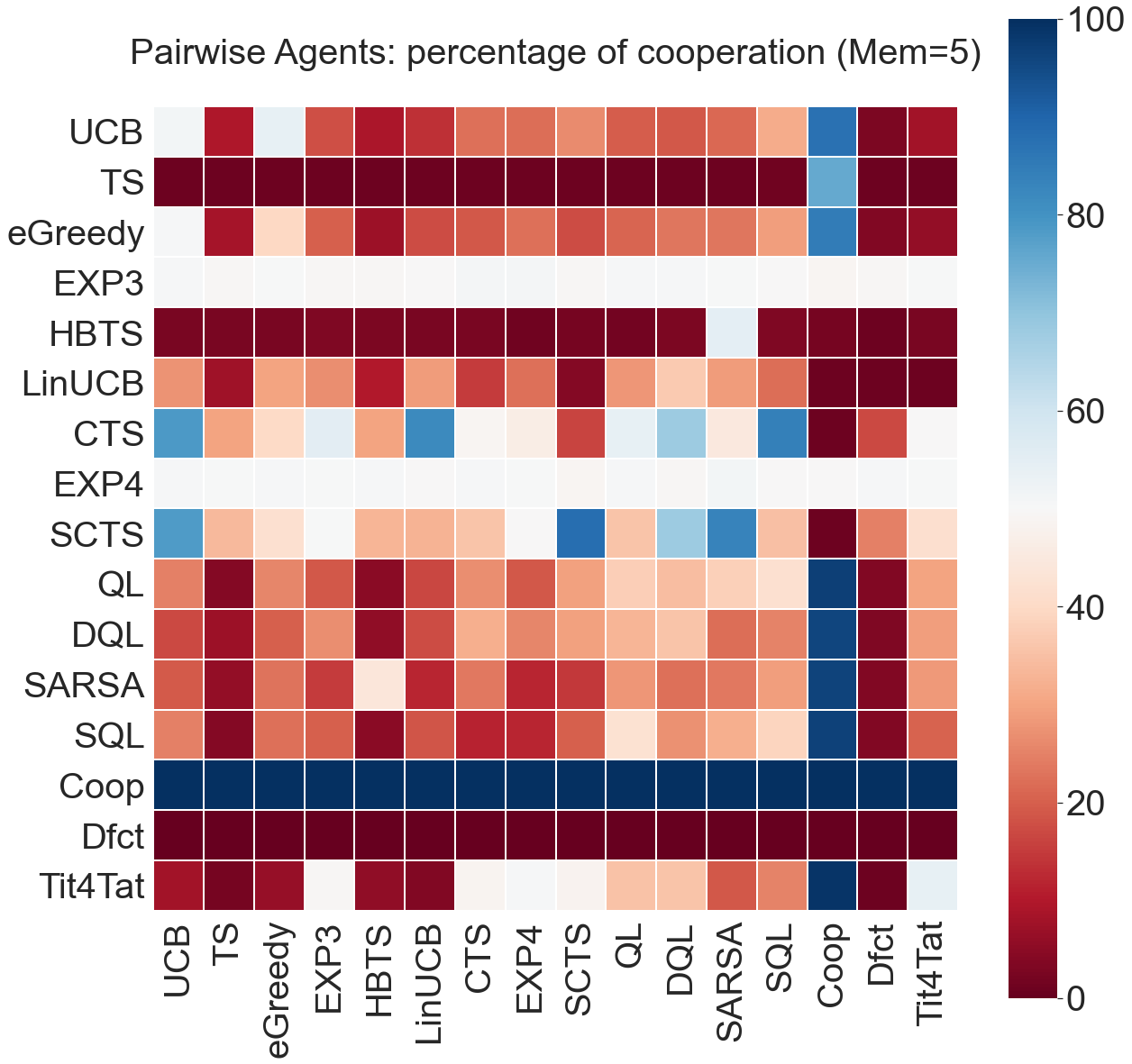}\hfill
\includegraphics[width=0.24\linewidth]{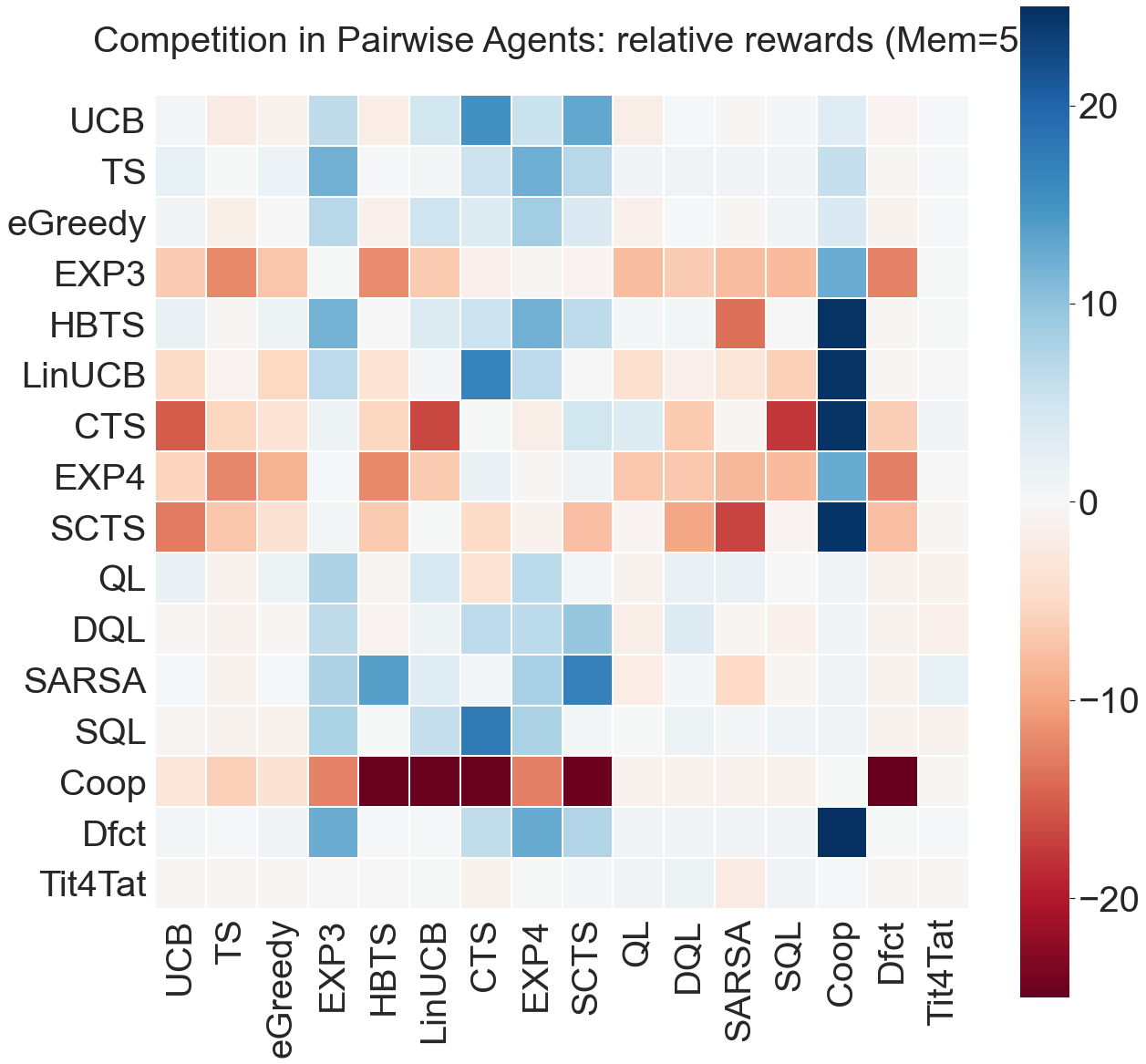}
% \par\caption{Cooperation and Competition in two-agent tournament: cooperation rate, relative rewards.}\label{fig:sd}
\par\caption{Success, Teamwork, Cooperation \& Competition in two-agent tournament.}\label{fig:rpsd}
\vspace{-1em}
\end{figure*}

Here, we describe the two main experiments we run, Iterated Prisoner's Dilemma (IPD) and Behavioral Cloning with Demonstration Rewards (BCDR).

% \begin{wraptable}{R}{0.25\textwidth}
\begin{table}[t]
% \begin{table}[tb]
\centering
 \vspace{-0.5em}
\caption{IPD Payoff}
\label{tab:ipd}
     
\resizebox{0.16\textwidth}{!}{
\centering
 \begin{tabular}{ l | c | c | }
\small
  & C & D \\ \hline
C  & R,R & S,T \\ \hline
D  & T,S & P,P \\ \hline
 \end{tabular}
 }
\vspace{-0.5em}
 \end{table}

\subsection{Iterated Prisoner's Dilemma (IPD)}

The Iterated Prisoner's Dilemma (IPD) can be defined as a matrix game $G = [N, \{A_i\}_{i\in N}, \{R_i\}_{i\in N}]$, where $N$ is the set of agents, $A_i$ is the set of actions available to agent $i$ with $\mathcal{A}$ being the joint action space $A_1 \times \cdots \times A_n$, and $R_i$ is the reward function for agent $i$. A special case of this generic multi-agent IPD is the classical two-agent case (Table \ref{tab:ipd}). In this game, each agent has two actions: cooperate (C) and defect (D), and can receive one of the four possible rewards: R (Reward), P (Penalty), S (Sucker), and T (Temptation). In the multi-agent setting, if all agents Cooperates (C), they all receive Reward (R); if all agents defects (D), they all receive Penalty (P); if some agents Cooperate (C) and some Defect (D), cooperators receive Sucker (S) and defector receive Temptation (T). The four payoffs satisfy the following inequalities: $T > R > P > S$ and $2R > T + S$. The PD is a one round game, but is commonly studied in a manner where the prior outcomes matter to understand the evolution of cooperative behaviour from complex dynamics \cite{axelrod1981evolution}.

\subsection{Behavioral Cloning with Demonstration Rewards}

Here we define a new type of multi-agent online learning setting, the Behavior Cloning with Demonstration Rewards (BCDR), present a novel training procedure and agent for solving this problem. In this setting, and similar to \cite{balakrishnan2019using,BalakrishnanBMR19,NoothigattuBMCM19} the agent first goes through a constraint learning phase where it is allowed to query the actions and receive feedback $r_k^e (t) \in [0, 1]$ about whether or not the chosen decision matches the teacher's action (from demonstration). During the deployment (testing) phase, the goal of the agent is to maximize both $r_k (t) \in [0, 1]$, the reward of the action $k$ at time $t$, and the (unobserved) $r_k^e (t) \in [0, 1]$, which models whether or not the taking action $k$ matches which action the teacher would have taken. During the deployment phase, the agent receives no feedback on the value of $r_k^e (t)$, where we would like to observe how the behavior captures the teacher's policy profile. In our specific problem, the human data plays the role of the teacher, and the behavioral cloning aims to train our agents to mimic the human behaviors.

%as the method section behind uses a similar idea.
     
\subsection{Online Learning Agents}

We briefly outlined the different types of online learning algorithms we have used: 

\textbf{Multi-Armed Bandit (MAB):} The multi-armed bandit algorithm models a sequential decision-making process, where at each time point a the algorithm selects an action from a given finite set of possible actions, attempting to maximize the cumulative reward over time \cite{LR85,UCB,Survey2019}. In the multi-armed bandit agent pool, we have Thompson Sampling (TS) \cite{T33}, Upper Confidence Bound (UCB) \cite{UCB}, epsilon Greedy (eGreedy) \cite{Sutton1998}, EXP3 \cite{auer2002nonstochastic} and the Human Based Thompson Sampling (HBTS) \cite{bouneffouf2017bandit}. 
 
\textbf{Contextual Bandit (CB).}
Following \cite{langford2008epoch}, this problem is defined as follows. At each time point (iteration), an agent is presented with a {\em context} ({\em feature vector}) before choosing an arm. In the contextual bandit agent pool, we have Contextual Thompson Sampling (CTS) \cite{AgrawalG13}, LinUCB \cite{LiCLW11}, EXP4 \cite{beygelzimer2011contextual} and Split Contextual Thompson Sampling (SCTS) \cite{lin2020unified,lin2021models}.

\textbf{Reinforcement Learning (RL).}
Reinforcement learning defines a class of algorithms for solving problems modeled as Markov decision processes (MDP) \cite{Sutton1998}. An MDP is defined by the tuple with a set of possible states, a set of actions and a transition function. In the reinforcement learning agent pool, we have Q-Learning (QL), Double Q-Learning (DQL) \cite{hasselt2010double}, State–action–reward–state–action (SARSA) \cite{rummery1994line} and Split Q-Learning (SQL) \cite{lin2019split,lin2020astory}. We also selected three most popular handcrafted policy for Iterated Prisoner's Dilemma: ``Coop'' stands for always cooperating, ``Dfct'' stands for always defecting and ``Tit4Tat'' stands for following what the opponent chose for the last time (which was the winner approach in the 1980 IPD tournament \cite{axelrod1980effective}).  

The choices of the agents evaluated in this work are the most common online learning agents in bandits, contextual bandits and reinforcement learning (the three online learning classes). We thought that competing them against one another, and competing the three online learning classes against one another might be an interesting experiment to study how the inductive bias of different methods yield different behaviors in social agent settings (e.g. iterated prisoners' dilemma).

\section{Results: Algorithms' Tournament}

\begin{figure*}[tb]

\centering
\includegraphics[width=0.23\linewidth]{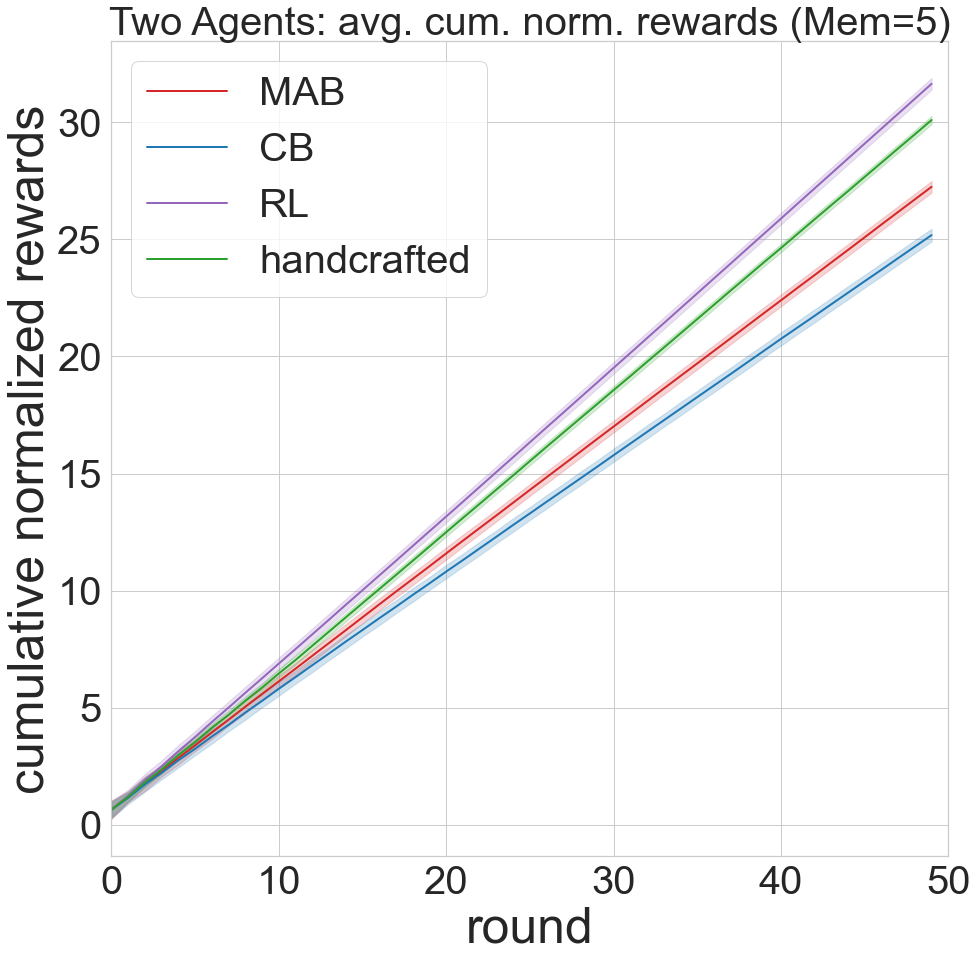} \hfill
\includegraphics[width=0.23\linewidth]{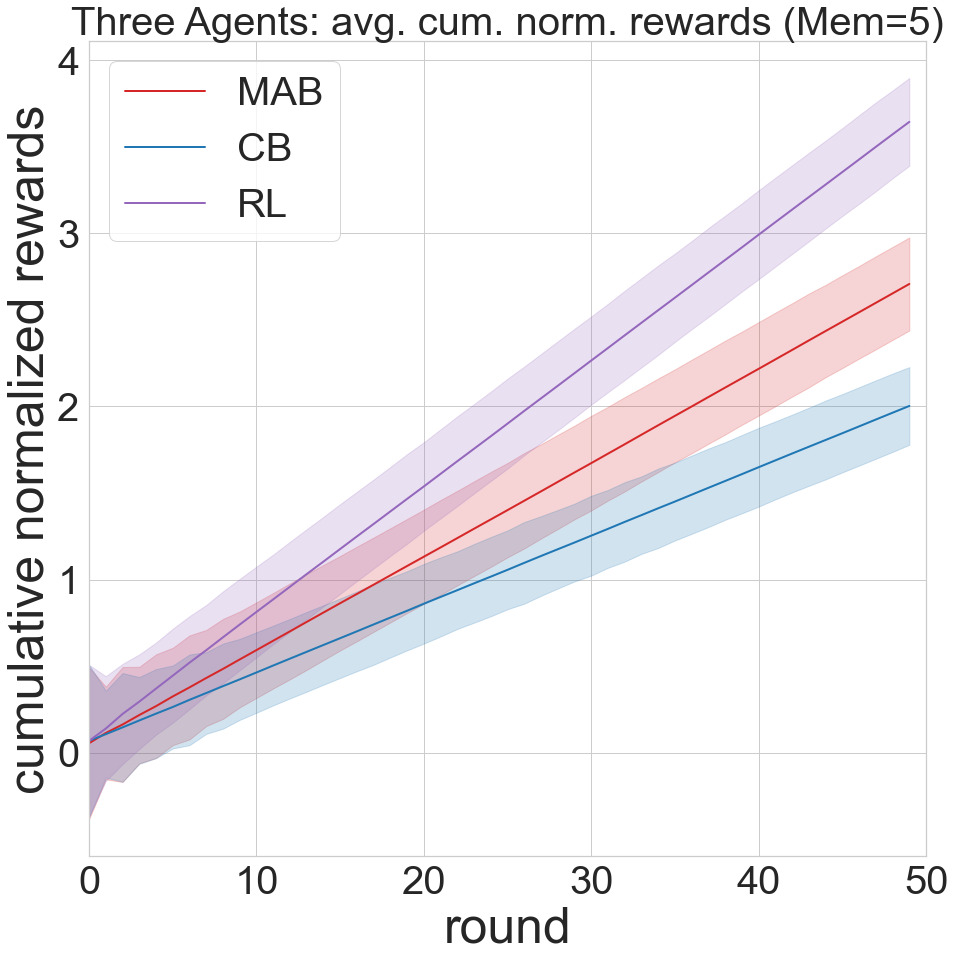} \hfill
\includegraphics[width=0.23\linewidth]{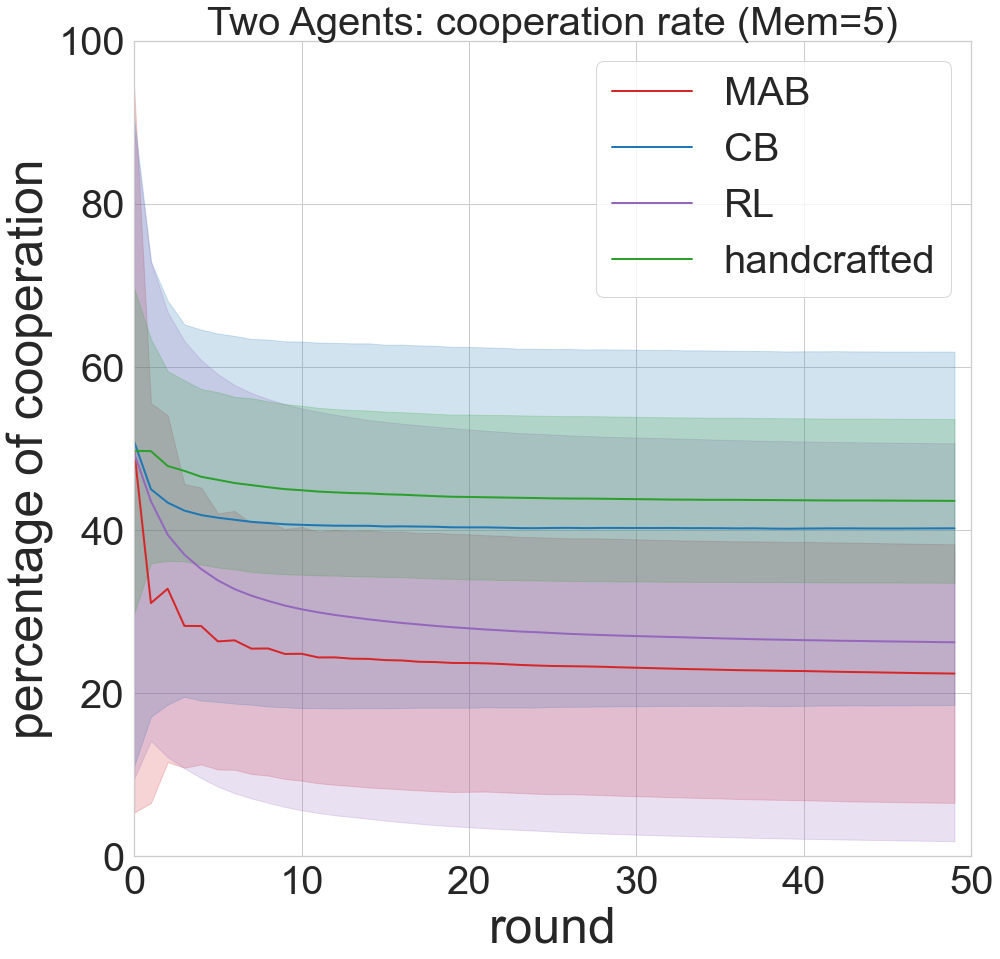} \hfill
\includegraphics[width=0.23\linewidth]{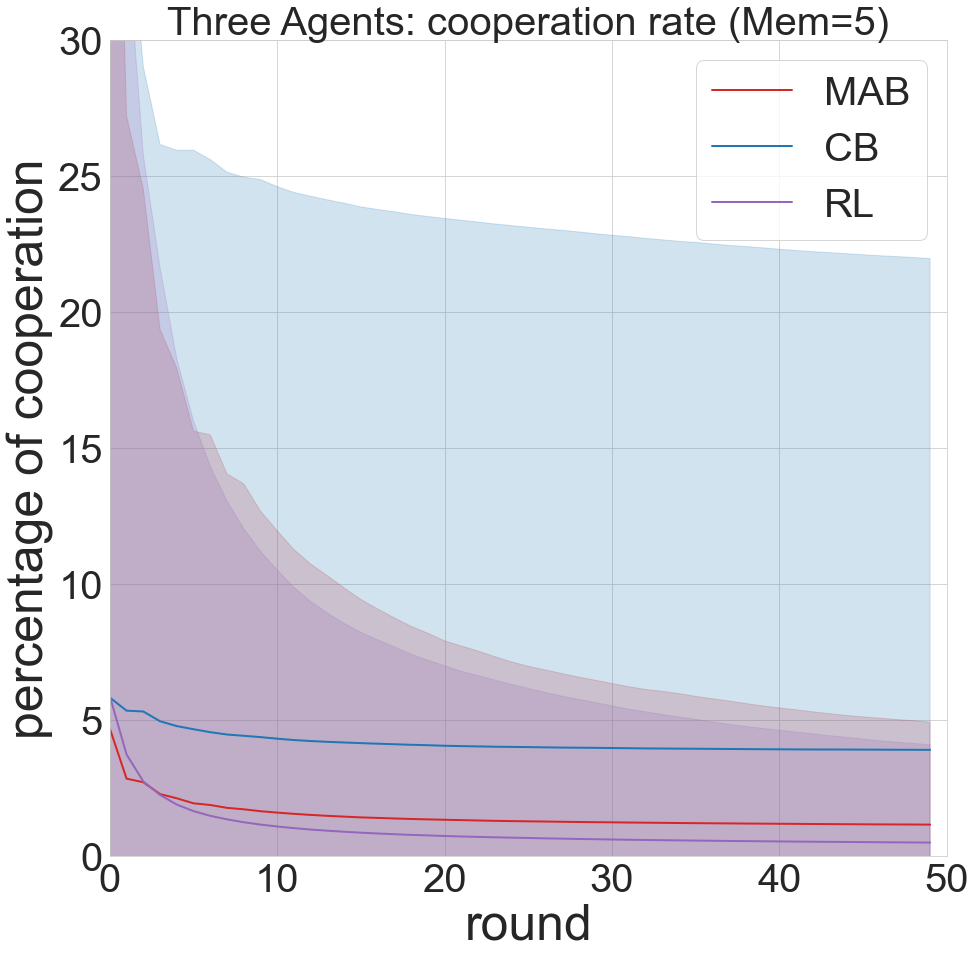} 

\caption{Cumulative reward and cooperation rate averaged by class in two- and three-player setting.}\label{fig:two_three}

\end{figure*}

\begin{figure*}[tb]
\centering
\includegraphics[width=0.18\linewidth]{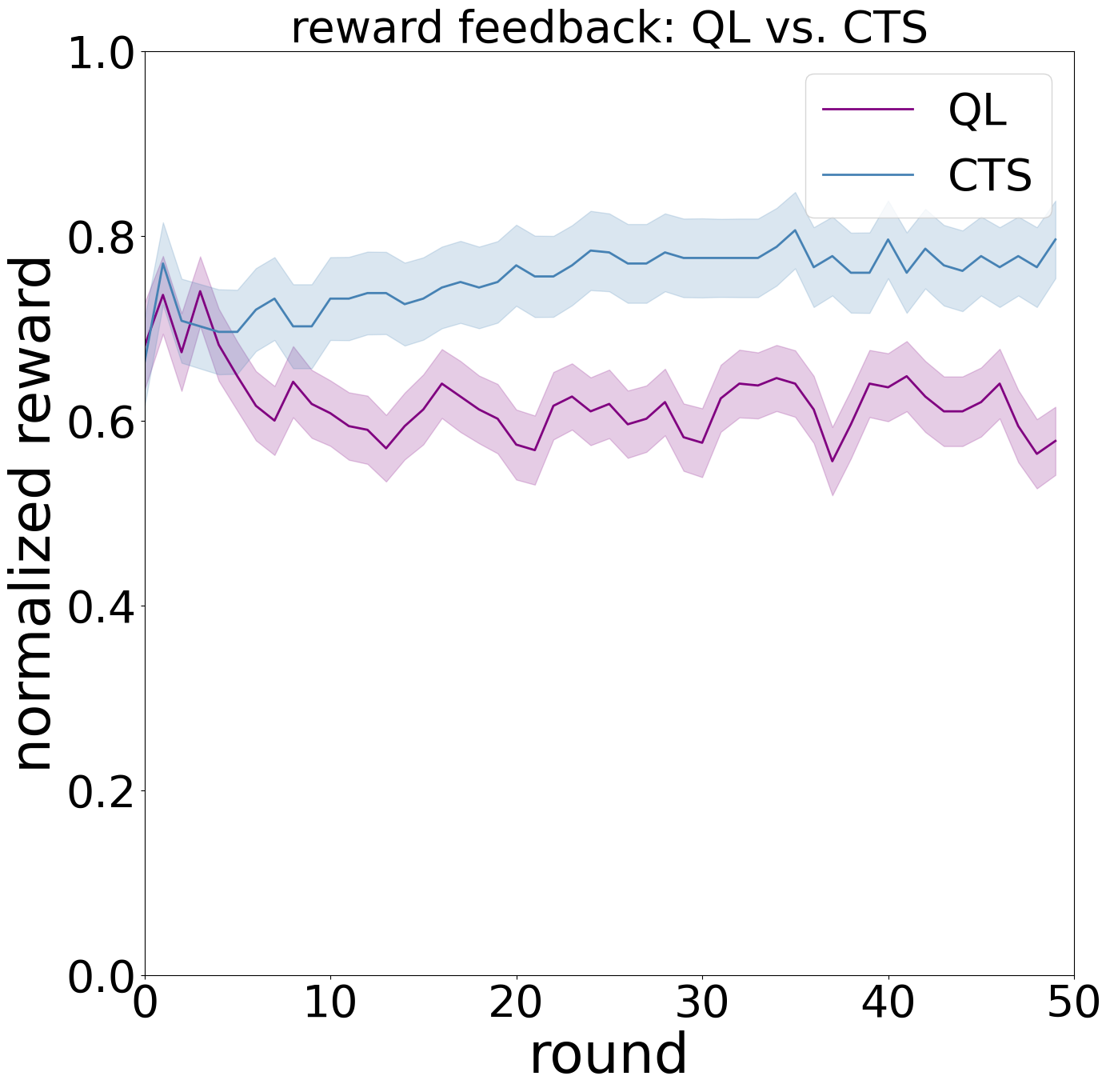} \hfill
\includegraphics[width=0.18\linewidth]{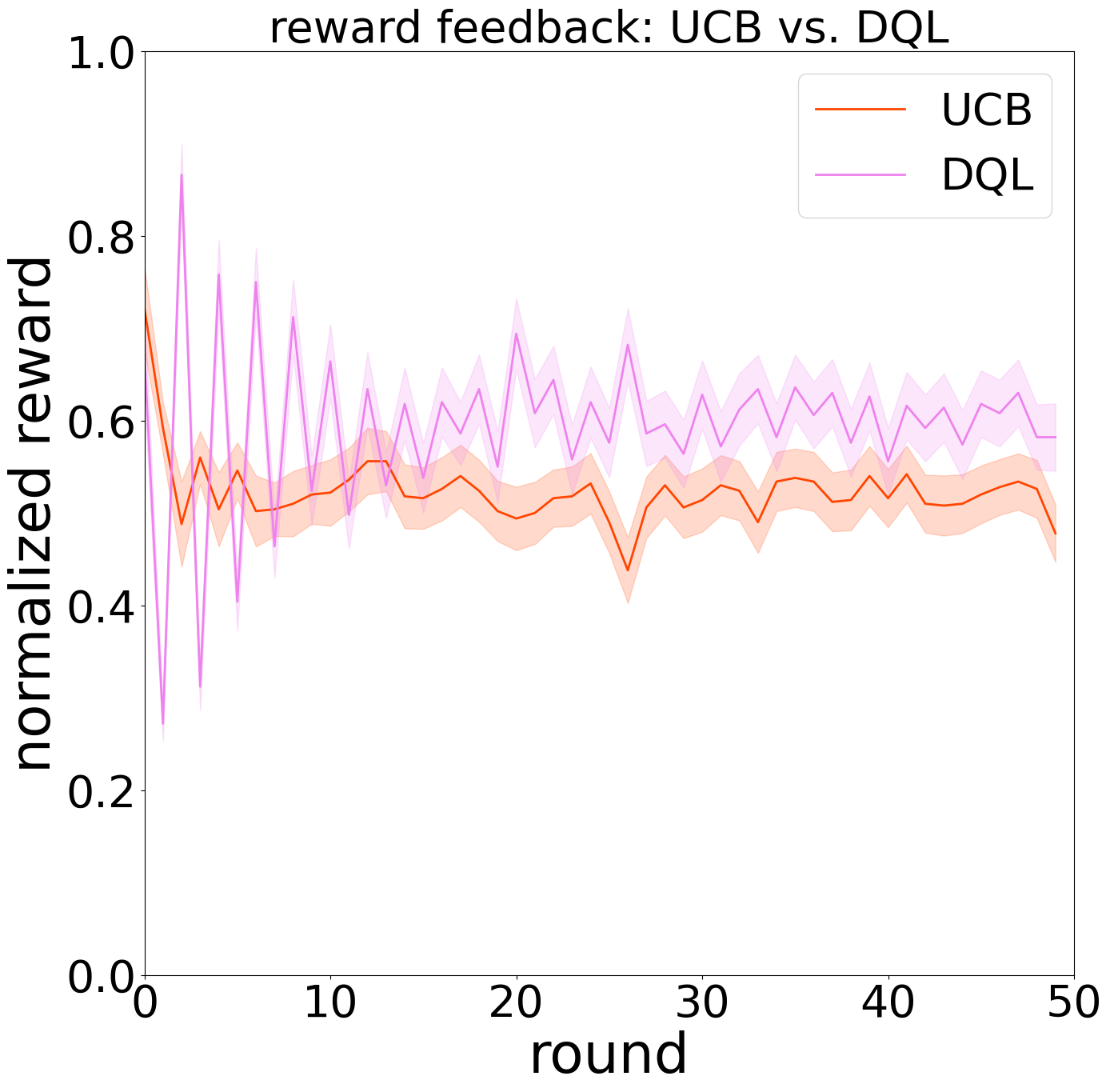} \hfill
\includegraphics[width=0.18\linewidth]{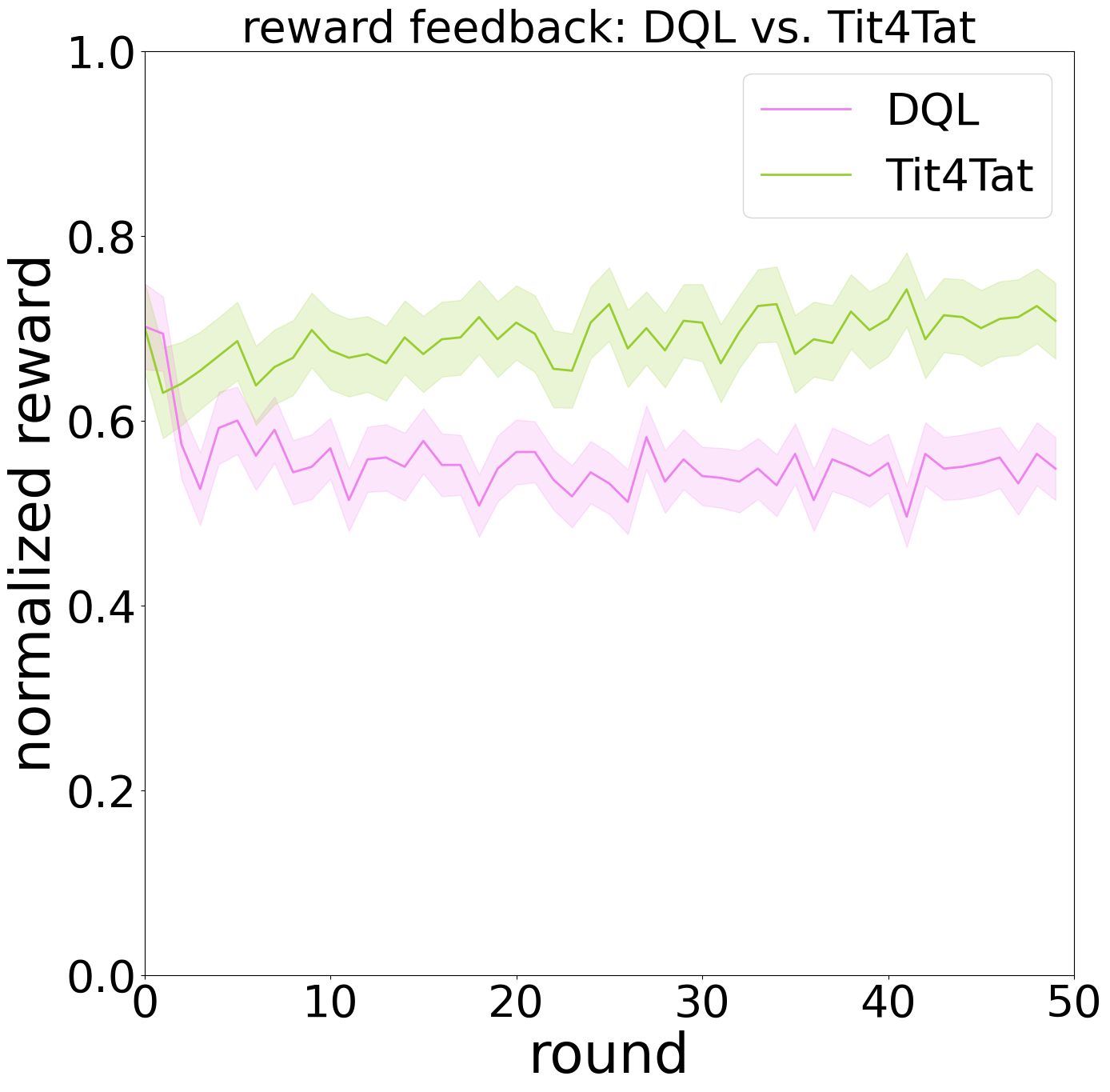} \hfill
\includegraphics[width=0.18\linewidth]{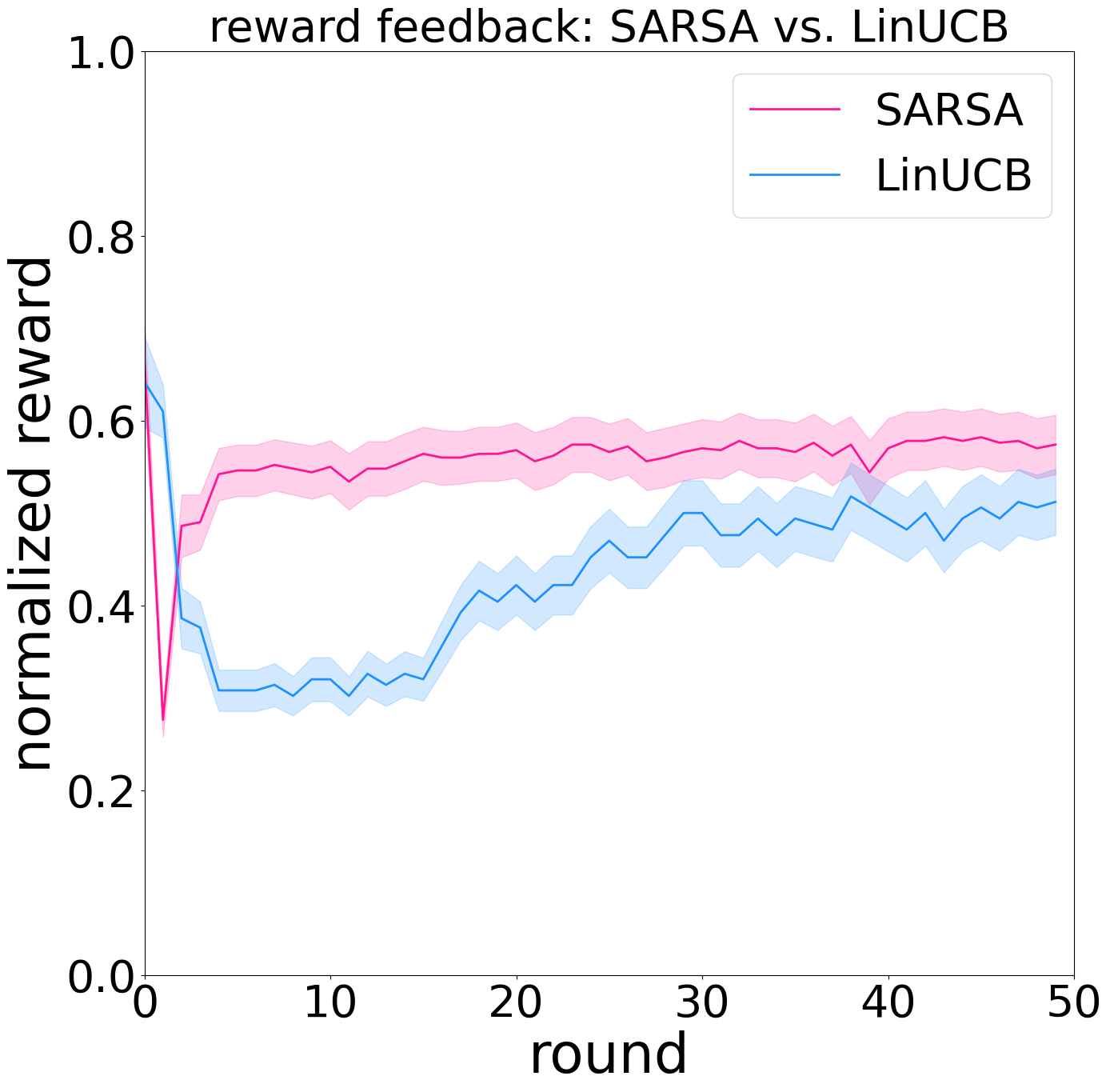} \hfill
\includegraphics[width=0.18\linewidth]{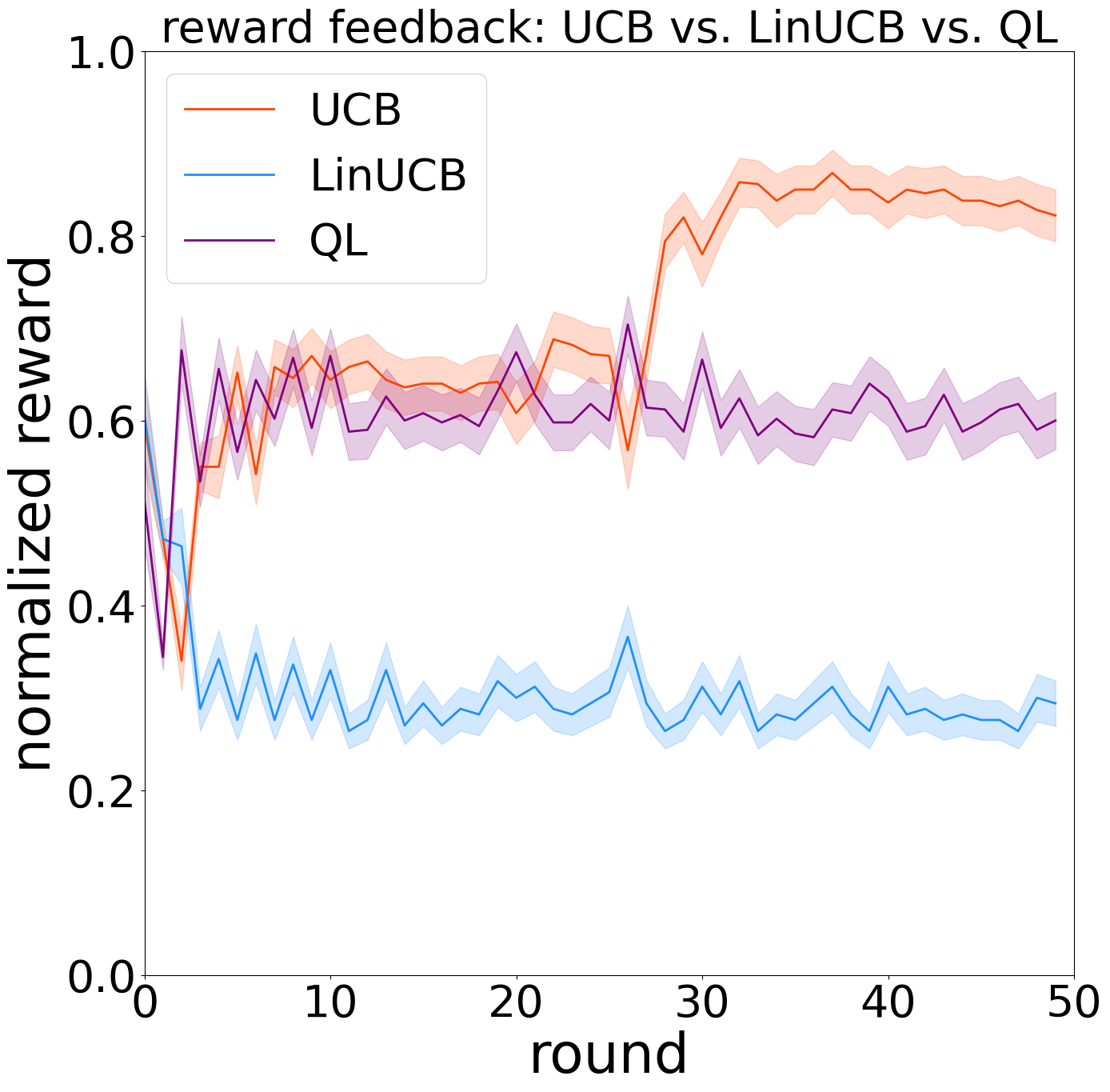}
\includegraphics[width=0.18\linewidth]{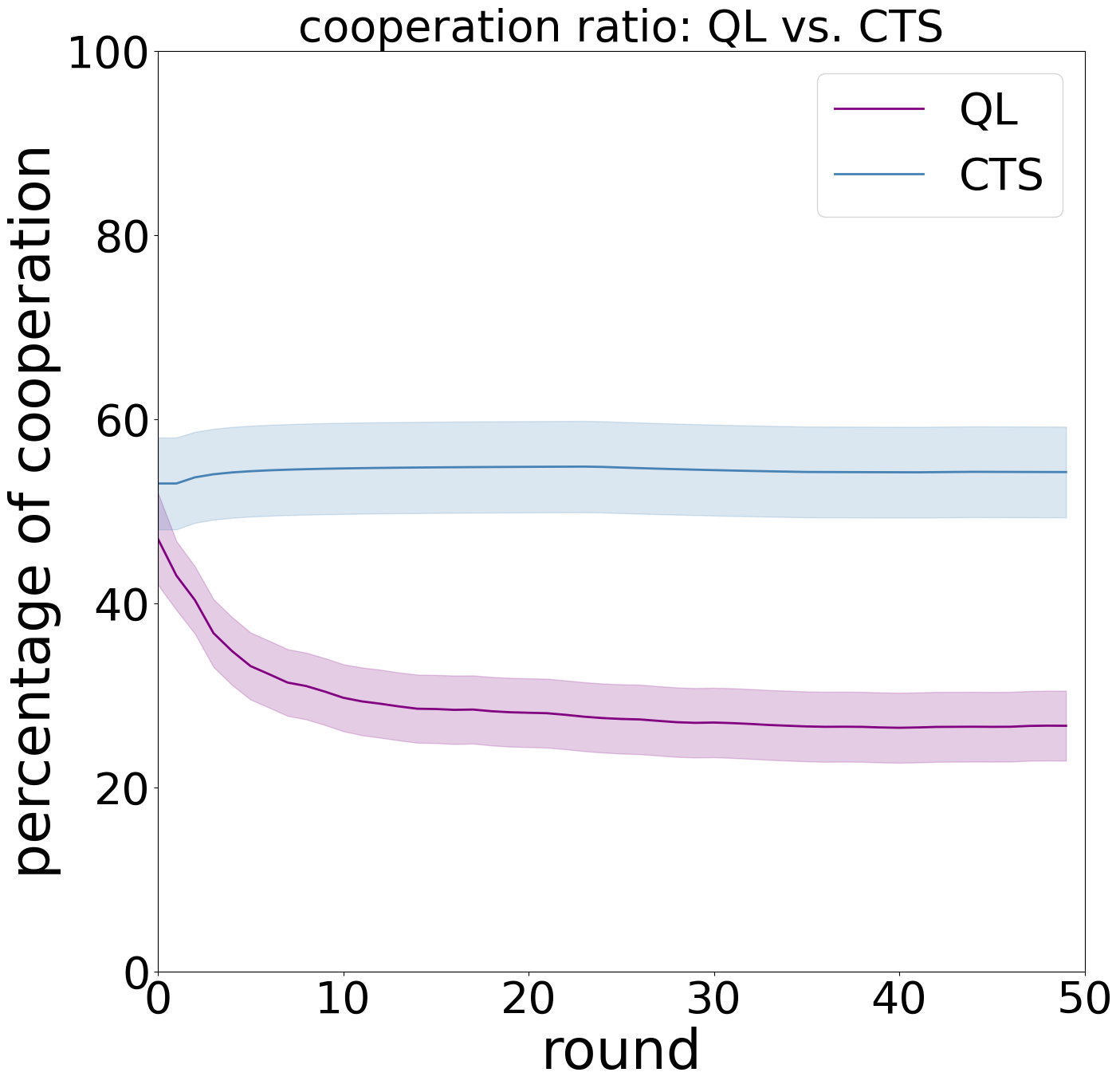} \hfill
\includegraphics[width=0.18\linewidth]{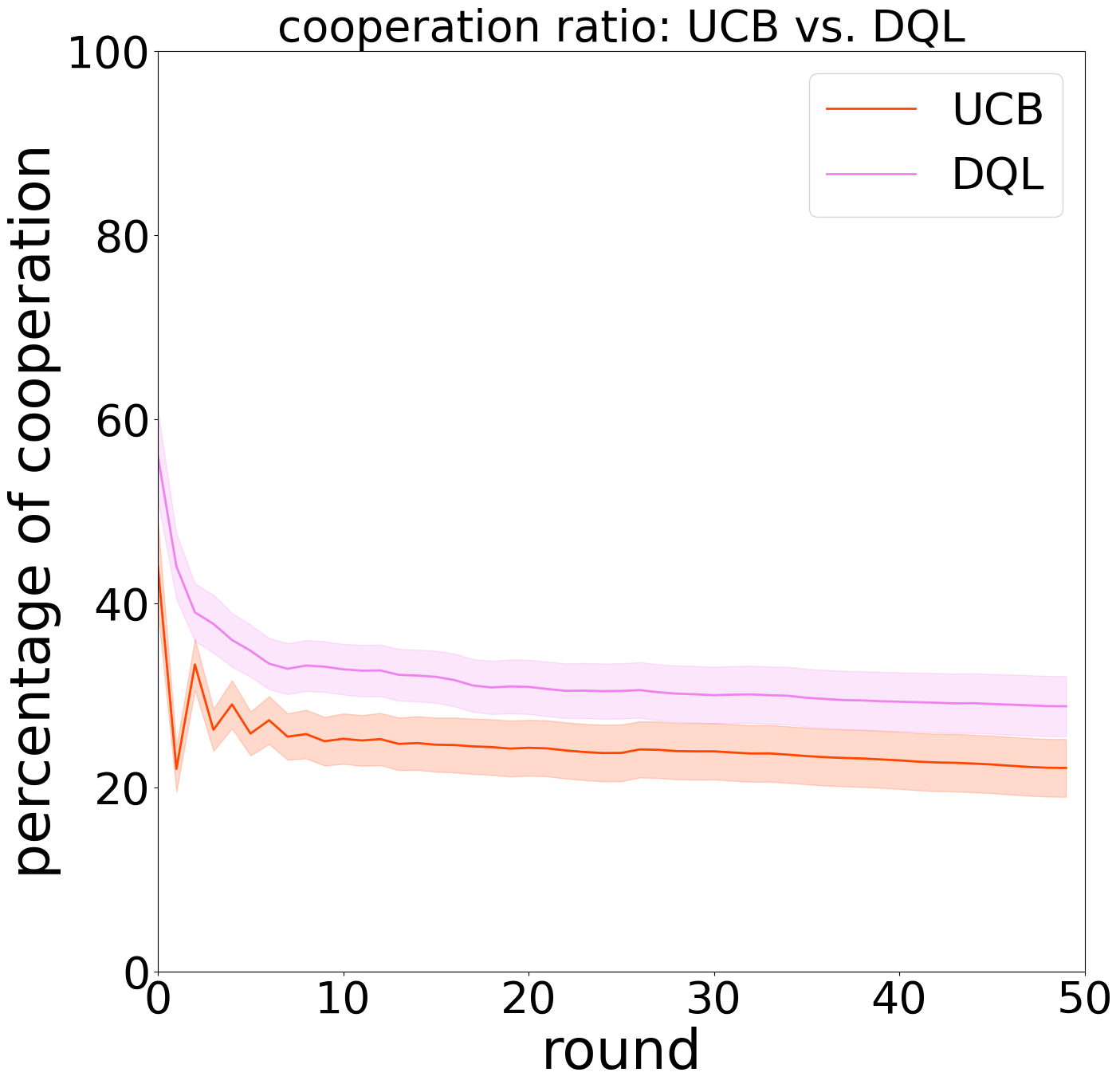} \hfill
\includegraphics[width=0.18\linewidth]{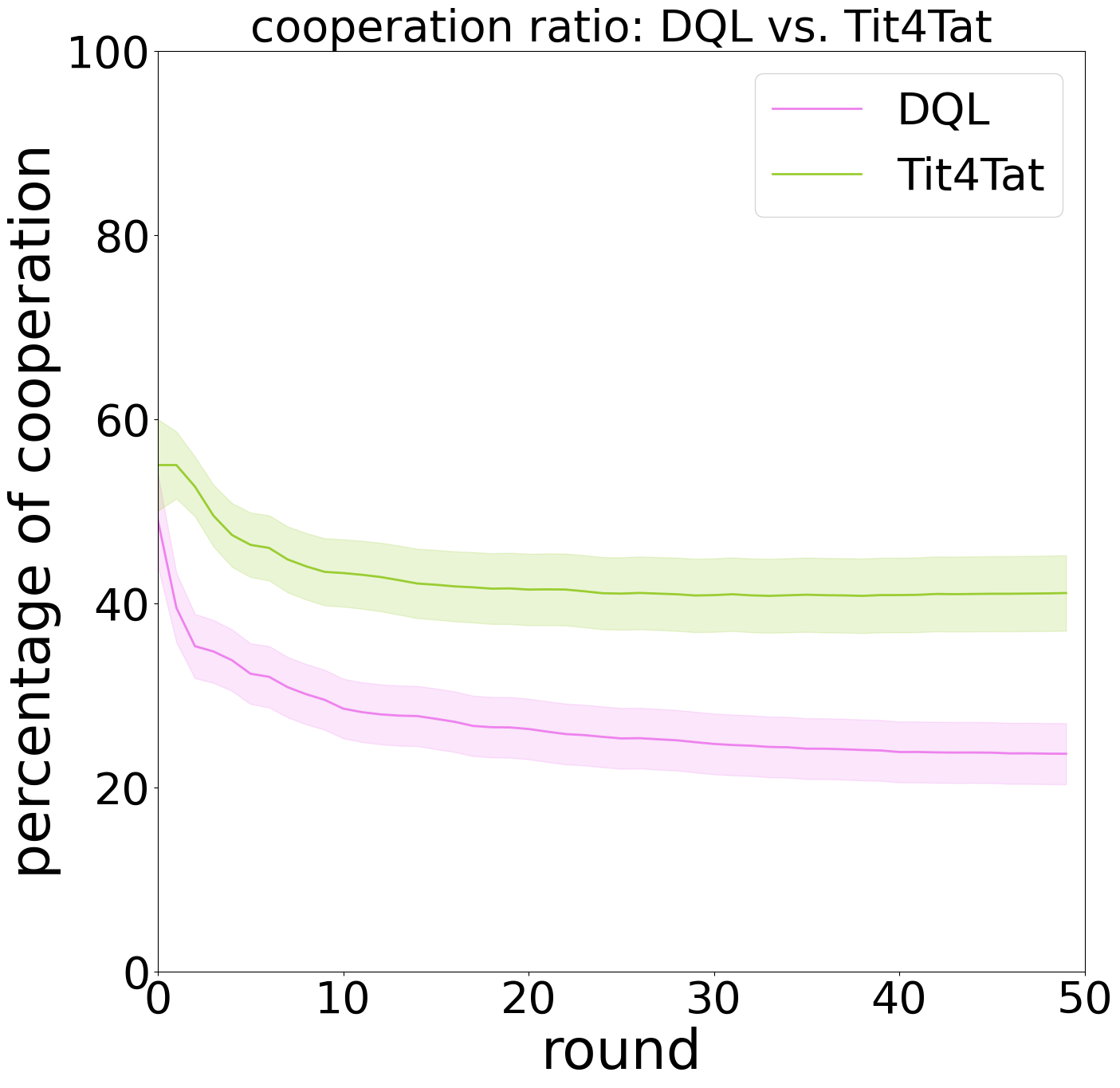} \hfill
\includegraphics[width=0.18\linewidth]{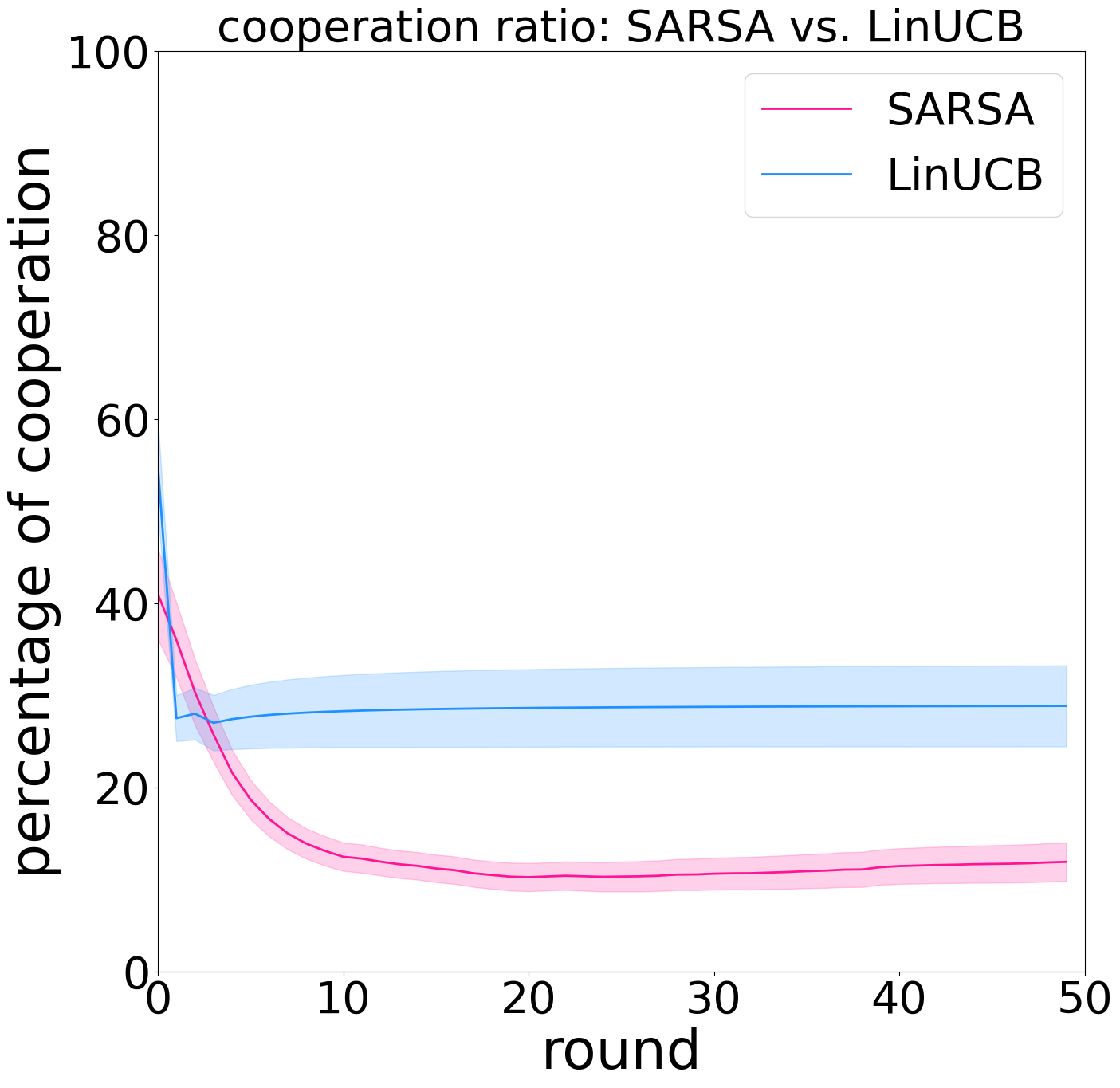} \hfill
\includegraphics[width=0.18\linewidth]{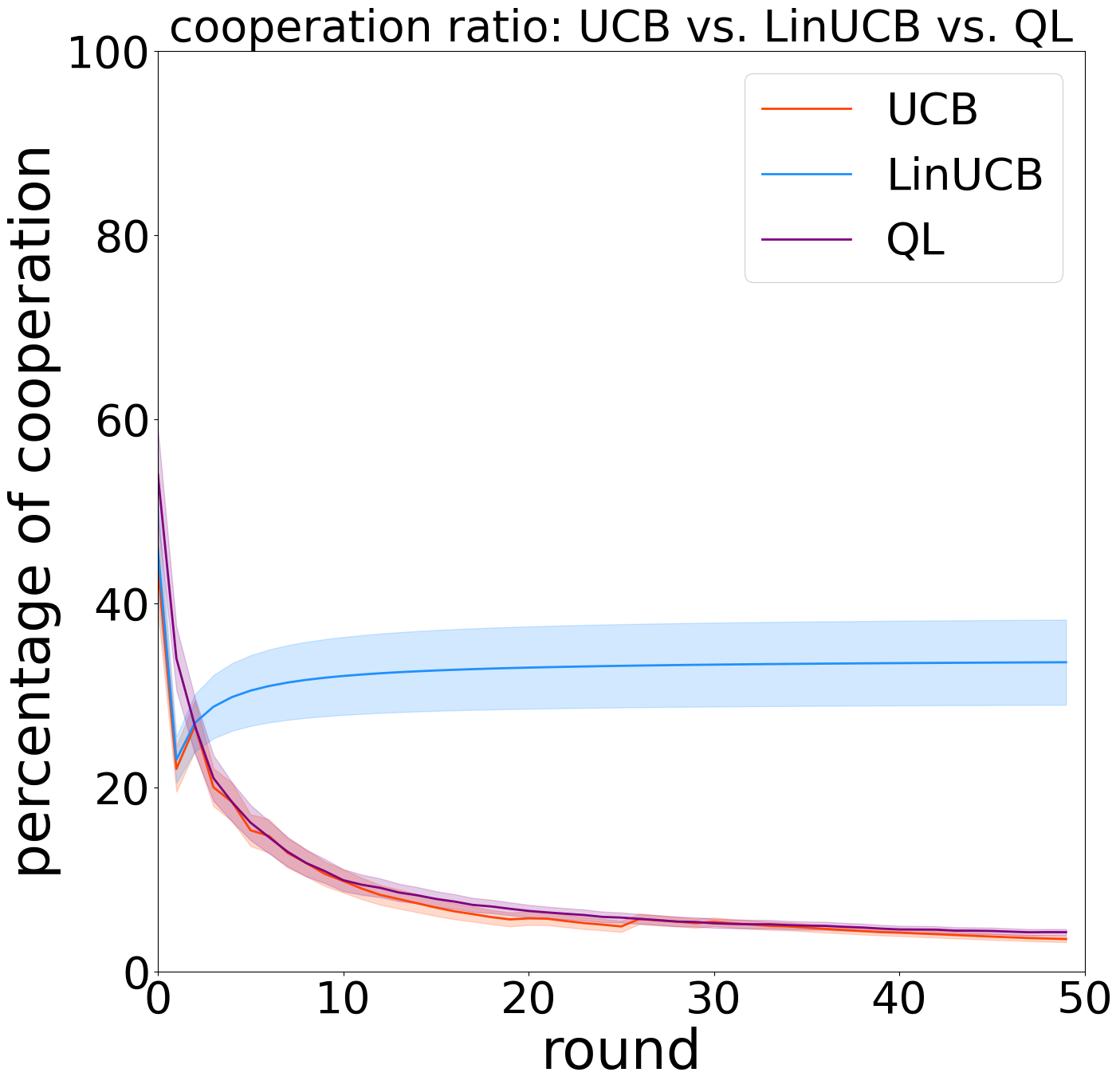}
\par\caption{Reward feedbacks and cooperation rates in some two-player and the three-player settings.}\label{fig:example}

\end{figure*}

\begin{figure*}[tb]
\centering
\includegraphics[width=0.45\linewidth]{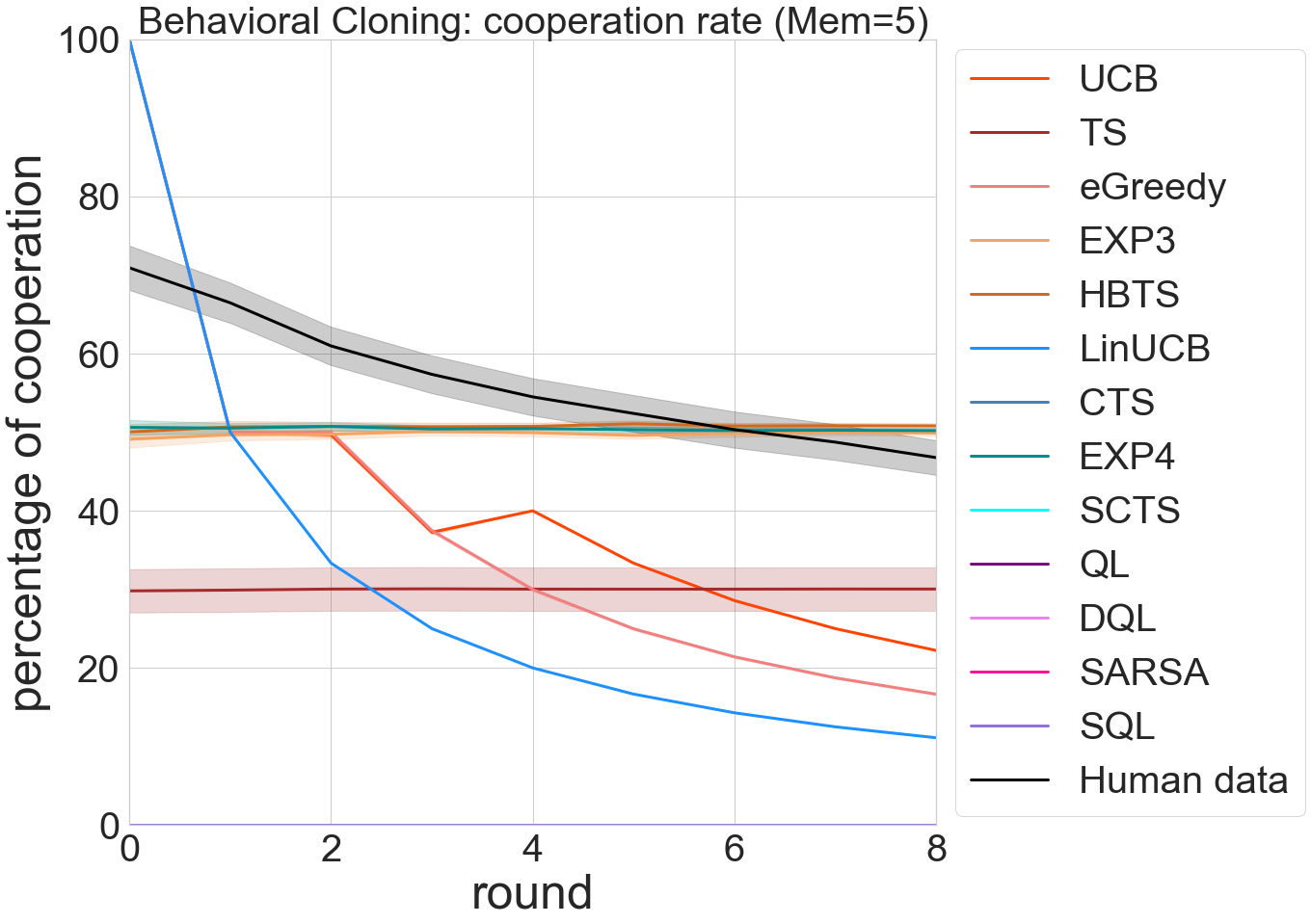} 
\hfill
\includegraphics[width=0.45\linewidth]{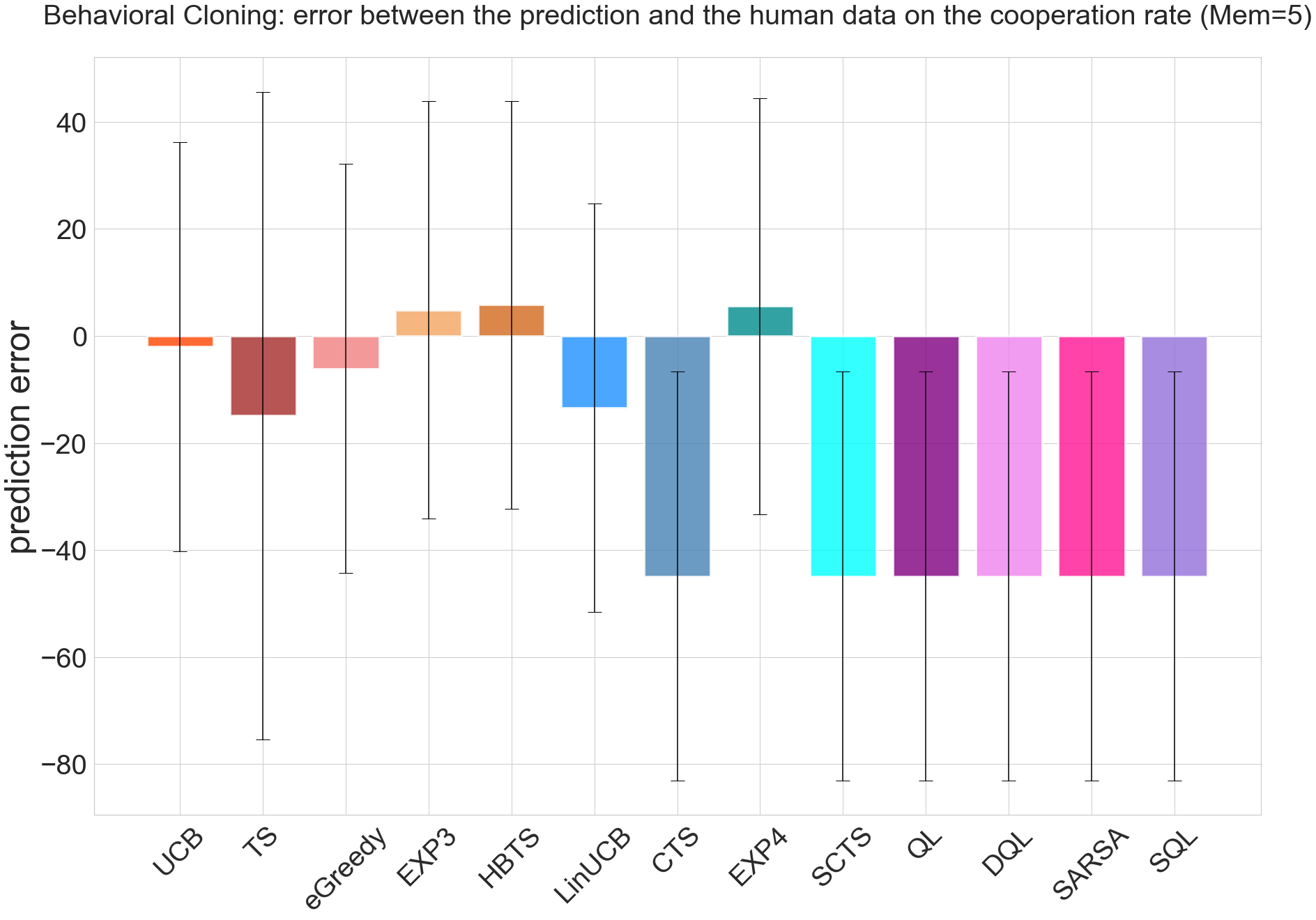}
\par\caption{Behavioral Cloning: bandits modeled human data the best with the lowest prediction error.}\label{fig:bclone}
\vspace{-1em}
\end{figure*}

\textbf{Game settings.} 
The payoffs are set as the classical IPD game: $T=5, R=3, P=1, S=0$. Following \cite{rapoport1965prisoner}, we create create standardized payoff measures from the R, S, T, P values using two differences
between payoffs associate with important game outcomes, both normalized by the difference
between the temptation to defect and being a sucker when cooperating as the other defects.
 
\textbf{State representations.} 
In most IPD literature, the state is defined the pair of previous actions of self and opponent. Studies suggest that only one single previous state is needed to define any prisoner’s dilemma strategy \cite{press2012iterated}. However, as we are interested in understanding the role of three levels of information (no information, with context but without state, and with both context and state), we expand the state representation to account for the past $n$ pairs of actions as the history (or memory) for the agents. For CB algorithms, this history is their context. For RL algorithms, this history is their state representation. In the following sections, we will present the results in which the memory is set to be the past 5 action pairs.
% (denoted $Mem=5$).
% ; $Mem=1$ in Appendix).
 
\textbf{Learning settings.} 
In all experiments, the discount factor $\gamma$ was set to be 0.95. The exploration is included with $\epsilon$-greedy algorithm with $\epsilon$ set to be 0.05 (except for the algorithms that already have an exploration mechanism). The learning rate was polynomial $\alpha_t(s, a) = 1/n_t(s, a)^{0.8}$, which was shown in previous work to be better in theory and in practice \cite{even2003learning}. All experiments were performed and averaged for at least 100 runs, and over 50 steps of dueling actions from the initial state. 
 
\textbf{Reported measures.} 
To capture the behavior of the algorithms, we report five measures: individual normalized rewards, collective normalized rewards, difference of normalized rewards, the cooperation rate and normalized reward feedback at each round. We are interested in the individual rewards since that is what online learning agents should effectively maximize their expected cumulative discounted reward for. We are interested in the collective rewards because it might offer important insights on the teamwork of the participating agents. We are interested in the difference between each individual player's reward and the average reward of all participating players because it might capture the internal competition within a team. We record the cooperation rate as the percentage of cooperating in all rounds since it is not only a probe for the emergence of strategies, but also the standard measure in behavioral modeling to compare human data and models \cite{nay2016predicting,lin2022predicting,lin2022predictinglstm}. Lastly, we provided reward feedback at each round as a diagnostic tool to understand the specific strategy emerged from each game. (The color codes throughout this paper are set constant for each of the 14 agents, such that all handcrafted agents have green-ish colors, MAB agents red-ish, CB agents blue-ish and RL agents purple-ish).

\subsection{Multi-Agent Tournament}

\textbf{Results for two-agent tournament. }
We record the behaviors of the agents playing against each other (and with themselves). Figure \ref{fig:rpsd} summarizes the reward and behavior patterns of the tournament. We first notice that MAB and RL algorithms learn to cooperate when their opponent is \textit{Coop}, yielding a high mutual reward, while CB algorithms mostly decide to defect on \textit{Coop} to exploit its trust. From the cooperation heatmap, we also observe that RL algorithms appear to be more defective when facing an MAB or CB algorithm than facing another RL algorithm. MAB algorithms are more defective when facing a CB algorithm than facing an RL or another MAB algorithm. Adversarial algorithms EXP3 and EXP4 fail to learn any distinctive policy. We also note interesting teamwork and competition behaviors in the heatmaps of collective rewards and relative rewards: CB algorithms are the best team players, yielding an overall highest collective rewards, followed by RL; RL are the most competitive opponents, yielding an overall highest relative rewards, followed by MAB.

Figure \ref{fig:two_three} summarizes the averaged reward and cooperation for each of the three classes, where we observe handcrafted algorithms the best, followed by RL algorithms and then MAB algorithms. CB algorithms receive the lowest final rewards among the four agent classes. Surprisingly, it also suggests that a lower cooperation rate don't imply a higher reward. The most cooperative learning algorithm class is CB, followed by RL. MAB, the most defective agents, don't score the highest. 

Detailed probing into specific games (Figure \ref{fig:example}) uncovers more diverse strategies than these revealed by the cooperation rates. For instance, in the game of QL vs. CTS, we observe that CTS converges to a fixed cooperation rate within the first few rounds and stayed constant since then, while the QL gradually decays its cooperation rate. In the game of UCB1 vs. DQL, UCB1 seemed to oscillate between a high and low cooperation rate within the first few rounds (because it is built to explore all actions first), while DQL gradually decays its cooperation rate. In DQL vs. Tit4Tat, we observe a seemingly mimicking effect of DQL to a tit-for-tat-like behaviors. In the game of SARSA vs. LinUCB, LinUCB converges to a fixed cooperation rate with the first few rounds and stays constant since then, while SARSA slowly decays its cooperation rate. There seems to be a universality of the three classes within the first few rounds. 

%\textbf{Technical interpretation of the numerical results. }In these analyses, we suggest the following reasons why RL algorithms are the best and CB algorithms are the worst in the two-agent IPD: (1) major learning events happen in the first few rounds; (2) the historical information are the most distinct in the first few rounds (as they might have not been defined when there are no history from say, five steps in the past); (3) the history is set as a sliding window; (3) CB algorithms treat this history as a linear mapping, so a shifting transform from the sliding window operation can be hurtful for its learning; (4) RL algorithms adopt a summing step from the feature spaces (for instance, in linear case, the Q value can be a weighted sum of the features), which stabilized the learning. 

\textbf{Cognitive interpretations of these learning systems. }
The main distinctions between the three classes of algorithms are the complexity of the learning mechanism and the cognitive system they adopt. In MAB setting, there is no attention to any contexts, and the agents aim to most efficiently allocate a fixed limited set of cognitive resources between competing (alternative) choices in a way that maximizes their expected gain. In CB setting, the agents apply an attention mechanism to the current context, and aim to collect enough information about how the context vectors and rewards relate to each other, so that they can predict the next best action to play by looking at the feature vectors. In RL setting, the agents not only pay attention to the current context, but also apply the attention mechanism to multiple contexts relate to different states, and aim to use the past experience to find out which actions lead to higher cumulative rewards. Our results suggest that in the Iterate Prisoner's Dilemma of two learning systems, an optimal learning policy should hold memory for different state representations and allocate attention to different contexts across the states, which explained the overall best performance by RL algorithms. This further suggests that in zero-sum games like the IPD, participating learning systems tend to undergo multiple states. The overall underperformance of CB suggests that the attention to only the current context was not sufficient without the state representation, because the learning system might mix the the context-dependent reward mappings of multiple states, which can oversimplify the policy and potentially mislead the learning as an interfering effect. On the other hand, MAB ignores the context information entirely, so they are not susceptible to the interfering effect from the representations of different contexts. Their learned policies, however, don't exhibit any interesting flexibility to account for any major change in the state (e.g., the opponent may just finish a major learning episode and switch strategies).

\textbf{Results for three-agent tournament. }
Here we wish to understand how all three classes of algorithms interact in the same arena. For each game, we pick one algorithm from each class (one from MAB, one from CB and one from RL) to make our player pool. We observe in Figure \ref{fig:two_three} a very similar pattern as the two-player case: RL agents demonstrate the best performance (highest final rewards) followed by MAB, and CB performed the worst. However, in three-agent setting, although CB is still the most cooperative, and RL became the most defective. More detailed probing into the specific games (Figure \ref{fig:example}) demonstrate more diverse strategies than these revealed by the cooperation rates. Take the game UCB1 vs. LinUCB vs. QL as an example, MAB algorithms start off as the most defective but later start to cooperate more in following rounds, while RL algorithms became more and more defective. CB in both cases stays cooperative at a relatively high rate.

\section{Behavioral Cloning with Human Data}

We collate the human data comprising 168,386 individual decisions from many human subjects experiments \cite{andreoni1993rational,bo2005cooperation,bereby2006speed}
% ,duffy2009cooperative,kunreuther2009bayesian,dal2011evolution,friedman2012continuous,fudenberg2012slow} 
that used real financial incentives and transparently conveyed the rules of the game to the subjects. As a a standard procedure in experimental economics, subjects anonymously interact with each other and their decisions to cooperate or defect at each time period of each interaction are recorded. They receive payoffs proportional to the outcomes in the same or similar payoff as the one we use in Table \ref{tab:ipd}. Following the similar preprocessing steps as \cite{nay2016predicting,lin2022predicting,lin2022predictinglstm}, we can construct the comprehensive collection of game structures and individual decisions from the description of the experiments in the published papers and the publicly available data sets. This comprehensive dataset consists of behavioral trajectories of different time horizons, ranging from 2 to 30 rounds, but most of these experimental data only host full historical information of at most past 9 actions. We further select only those trajectories with these full historical information, which comprise 8,257 behavioral trajectories. We randomly select 8,000 of them as training set and the other 257 as test set. 

In the training phase, all agents are trained with the demonstration rewards as feedback sequentially for the trajectories in the training set. In the testing phase, we paused all the learning, and tested on 257 trajectories independently, recorded their cooperation rate. In each test trajectory, we compared their evolution of cooperation rate to that of the human data and compute a prediction error.

Figure \ref{fig:bclone} summarizes the testing results of all the agents in predicting the actions and their cooperation rates from human data. From the heatmap of the cooperation rates, we observe that the behavioral policy that each agent cloned from the data varies by class. RL algorithms all seem to learn to defect at all costs (``tragedy of the commons''). CB algorithms mostly converge to a policy that adopted a fixed cooperation rate. Comparing with the other two, MAB algorithms learn a more diverse cooperation rates across test cases. The line plot on the right confirms our understanding.The cooperation rate by the real humans (the black curve) tends to decline slowly from around 70\% to around 40\%. UCB1 and epsilon Greedy both captured the decaying properties, mimicing the strategy of the human actions. The prediction error analysis matches this intuition. The UCB1 and epsilon greedy (or MAB algorithms in general), appear to be best capturing human cooperation. 
% \footnote{We would also like to point out the importance of the sanity check from the line plot (the cooperation rate vs. round). In the prediction error figures, EXP3 and EXP4 seem to have an overall low error, but this can be misleading: from the cooperation rate figures, we note that EXP3 and EXP4 don't seem to learn any policy at all (randomly choosing at 50\% over the entire time), while the other agents all appear to adopte a non-random strategy.}

\section{Clinical Evidences and Implications}

Evidence has linked dopamine function to reinforcement learning via midbrain neurons and connections to the basal ganglia, limbic regions, and cortex. Neuron firing rates computationally represent reward magnitude, expectancy, and violations (prediction error) and other value-based signals \cite{Schultz1997}, allowing an animal to update and maintain value expectations associated with particular states and actions. When functioning properly, this helps an animal develop a policy to maximize outcomes by approaching/choosing cues with higher expected value and avoiding cues associated with loss or punishment. This is similar to reinforcement learning widely used in computing and robotics \cite{Sutton1998}, suggesting mechanistic overlap in humans and AI. Evidence of Q-learning and actor-critic models have been observed in spiking activity in midbrain dopamine neurons in primates \cite{Bayer2005} and in human striatum by blood-oxygen-level-dependent imaging (BOLD) \cite{ODoherty2004}. 

The literature on the reward processing abnormalities in particular neurological and psychiatric disorders is quite extensive; below we summarize some of the recent developments in this fast-growing field. It is well-known that the neuromodulator dopamine plays a key role in reinforcement learning processes. Parkinson's disease (PD) patients, who have depleted dopamine in the basal ganglia, tend to have impaired performance on tasks that require learning from trial and error. For example, \cite{frank2004carrot} demonstrate that off-medication PD patients are better at learning to avoid choices that lead to negative outcomes than they are at learning from positive outcomes, while dopamine medication typically used to treat PD symptoms reverses this bias. Alzheimer's disease (AD) is the most common cause of dementia in the elderly and, besides memory impairment, it is associated with a variable degree of executive function impairment and visuospatial impairment. As discussed in \cite{perry2015reward}, AD patients have decreased pursuit of rewarding behaviors, including loss of appetite; these changes are often secondary to apathy, associated with diminished reward system activity. Moveover, poor performance on certain tasks is associated with memory impairments. Frontotemporal dementia (bvFTD) usually involves a progressive change in personality and behavior including disinhibition, apathy, eating changes, repetitive or compulsive behaviors, and loss of empathy \cite{perry2015reward}, and it is hypothesized that those changes are associated with
abnormalities in reward processing. For instance, alterations in eating habits with a preference for carbohydrate sweet rich foods and overeating in bvFTD patients can be associated with abnormally increased reward representation for food, or impairment in the negative (punishment) signal associated with fullness.  Authors in \cite{luman2009does} suggest that the strength of the association between a stimulus and the corresponding response is more susceptible to degradation in  Attention-deficit/hyperactivity disorder (ADHD) patients, which suggests problems with storing the stimulus-response associations. Among other functions, storing the associations requires working memory capacity, which is often impaired in ADHD patients. \cite{redish2007reconciling} demonstrated that patients suffering from addictive behavior have heightened stimulus-response associations, resulting in enhanced reward-seeking behavior for the stimulus which generated such association. \cite{taylor2016mesolimbic} suggested that chronic pain can elicit in a hypodopaminergic (low dopamine) state that impairs motivated behavior, resulting into a reduced drive in chronic pain patients to pursue the rewards. Reduced reward response may underlie a key system mediating the anhedonia and depression, which are common in chronic pain. 
% A variety of computational models was proposed for studying the disorders of reward processing in specific disorders  \cite{frank2004carrot,seeley2012frontotemporal,hauser2016computational,dezfouli2009neurocomputational,redish2007reconciling,hess2014beyond}. 

\section{Discussion}

The broader motivation of this work is to increase the two-way traffic between artificial intelligence and neuropsychiatry, in the hope that a deeper understanding of brain mechanisms revealed by how they function (``neuro'') and dysfunction (``psychiatry'') can provide for better AI models, and conversely AI can help to conceptualize the otherwise bewildering complexity of the brain. 

The behavioral cloning results suggest that bandit algorithms (without context) are the best in term of fitting the human data, which open the hypothesis that human are not considering the context when they are playing the iterated prisoner’s dilemma. This discovery proposes new modeling effort on human study in the bandit framework, and points to future experimental designs which incorporate these new parametric settings and control conditions. In particular, we propose that our approach may be relevant to study reward processing in different mental disorders, for which some mechanistic insights are available. A body of recent literature has demonstrated that a spectrum of neurological and psychiatric disease symptoms are related to biases in learning from positive and negative feedback \cite{Maia2011}. Studies in humans have shown that when reward signaling in the direct pathway is over-expressed, this may enhance state value and incur pathological reward-seeking behavior, like gambling or substance use. Conversely, enhanced aversive error signals result in dampened reward experience thereby causing symptoms like apathy, social withdrawal, fatigue, and depression. Both genetic predispositions and experiences during critical periods of development can predispose an individual to learn from positive or negative outcomes, making them more or less at risk for brain-based illnesses \cite{Holmes2018}. This highlight our need to understand how intelligent systems learn from rewards and punishments, and how experience sampling may impact reinforcement learning during influential training periods. Simulation results of the mental variants matches many of the clinical implications presented here, but also points to other complications from the social setting that deserve future investigation.

The approach proposed in the present manuscript, we hope, will contribute to expand and deepen the dialogue between AI and neuropsychiatry.

\section{Conclusion}

In this work, we explore the full spectrum of online learning agents: multi-armed bandits, contextual bandits and reinforcement learning. To quantitatively study their behaviors, we evaluate them based on a series of tournaments of iterated prisoner's dilemma. This allows us to analyze the dynamics of policies learned by multiple self-interested independent reward driven agents, where we observe that the contextual bandit is not performing well in the tournament, which means that considering the current situation to make decision is the worst in this kind of game. Basically we should either not care about the current situation or caring about more situations, but not just the current one. We have also studied the capacity of these algorithms to fit the human behavior. We observed that bandit algorithms (without context) are the best in term of fitting the human data, which opens the hypothesis that human are not considering the context when they are playing the IPD. Next steps include extending our evaluations to other sequential social dilemma environments with more complicated and mixed incentive structure, such as fruit Gathering game and Wolfpack hunting game \cite{leibo2017multi,wang2018towards}, comparing these mechanistic decision making models with predictive modeling surrogate models \cite{lin2022predicting,lin2022predictinglstm}, and building reinforcement learning-based recommendation systems that model properties of human decision making \cite{lin2022super}.

% \clearpage
% \vspace{1em}
% \vspace{2em}
\bibliographystyle{splncs04}
{
% \scriptsize
\bibliography{main}
}
% \clearpage

% \input{./sec_appendix}

\end{document}